\documentclass{PoS}

\def\asec{\ifmmode ^{\prime\prime}\else$^{\prime\prime}$\fi}

\def\degs{\ifmmode ^{\circ}\else$^{\circ}$\fi}
\def\amin{\ifmmode ^{\prime}\else$^{\prime}$\fi}
\def\asec{\ifmmode ^{\prime\prime}\else$^{\prime\prime}$\fi}

\def\degs{\ifmmode ^{\circ}\else$^{\circ}$\fi}
\def\amin{\ifmmode ^{\prime}\else$^{\prime}$\fi}

\def\EE#1{\times 10^{#1}}

\def\cm{\mbox{\,cm}}

\def\cm3{\mbox{\,cm$^{-3}$}}

\def\ergshz{\mbox{~erg~s$^{-1}$~Hz$^{-1}$}}

\def\lsim{\!\!\!\phantom{\le}\smash{\buildrel{}\over
 {\lower2.5dd\hbox{$\buildrel{\lower2dd\hbox{$\displaystyle<$}}\over
                                 \sim$}}}\,\,}
\def\gsim{\!\!\!\phantom{\ge}\smash{\buildrel{}\over
{\lower2.5dd\hbox{$\buildrel{\lower2dd\hbox{$\displaystyle>$}}\over
                               \sim$}}}\,\,}

\title{Core collapse supernovae and starbursts}

\ShortTitle{Core collapse supernovae and starbursts}

\author{\speaker{Miguel A. P\'erez-Torres}\\
        Instituto de Astrofísica de Andalucía - CSIC, 18080 Granada, Spain\\
        E-mail: \email{torres@iaa.es} 
               }

               \abstract{Core-collapse supernovae (CCSNe; Type Ib/c
                 and Type II SNe) are the endproducts of massive stars
                 (M $\geq 8$ M$_{\odot}$), and yield radio events
                 whose brightness depends on the intensity of the
                 interaction experienced by the supernova ejecta with
                 the circumstellar presupernova wind material
                 \cite{chevalier82}.  The fact that CCSNe are
                 intrinsically radio supernovae --albeit with a huge
                 range of different radio powers-- and hence
                 unaffected by dust absorption, together with the high
                 resolution and high sensitivity provided by current
                 VLBI arrays, has been exploited to directly image the
                 radio brightness structure of CCSNe in nearby ($D
                 \lsim 20$ Mpc) galaxies.  This has allowed to gain
                 insight into the physics of both CCSNe and of the
                 circumstellar medium (CSM) with which they interact.
                 In addition, ultra-high-resolution,
                 ultra-high-sensitivity radio observations of CCSNe in
                 Luminous and Ultra-Luminous Infrared Galaxies (LIRGs
                 and ULIRGs, respectively) in the local Universe, can
                 be used to directly measure of the current CCSN rate
                 and star formation rate (SFR), given an assumed
                 initial mass function (IMF). \\

                 In this contribution, I give a brief overview of VLBI
                 observations made of some CCSNe in nearby galaxies,
                 and then present some of the most relevant results
                 obtained with high-resolution radio observations of
                 (U)LIRGs in the local Universe, aimed at directly
                 detecting CCSNe via their radio emission, and thus
                 determine their CCSN and star formation rates,
                 independently of models. This is of particular
                 relevance, in view of the fact that our estimates of
                 star formation (and CCSN) rates in high-z starburst
                 galaxies relies on standard relationships between
                 far-infrared luminosity and star-formation rate. In
                 particular, I will present recently obtained
                 results with the e-EVN on the nuclear region of Arp 299-A.}

\FullConference{Science and Technology of Long Baseline Real-Time Interferometry: \\
The 8th International e-VLBI Workshop, EXPReS09\\
		 June 22 - 26 2009\\
		 Madrid, Spain}

\begin{document}

\section{Circumstellar interaction and VLBI imaging of core-collapse
  supernovae in nearby galaxies}

Type II supernovae (SNe) are associated with massive stars that have
expelled slow, dense winds during their supergiant phase.  The stellar
explosion drives a shock into this wind, at speeds as high as 20000
${\rm km\,s^{-1}}$ and temperatures of $\sim\,10^9$\,K.  In addition,
a reverse shock propagates back into the stellar envelope at speeds of
500-1000 ${\rm km\,s^{-1}}$ relative to the expanding ejecta.  This is
the so-called standard interaction model \cite{chevalier82}, and
radio, optical, and X-ray emission from Type II supernovae have been
usually interpreted within this model.  The outgoing shock forms a
high-energy-density shell that is responsible for the production of
synchrotron radio emission, while the reverse shock accounts for the
optical and soft X-ray emission.

A beautiful confirmation that the above scenario applies to Type II
SNe came when SN 1993J (Type IIb) in M81 (D$\approx$3.63 Mpc) was
imaged with VLBI just six months after its explosion, showing a
shell-like radio structure \cite{marcaide95a}. Its first year of
expansion proceeded in a self-similar way \cite{marcaide95b}, in
perfect agreement with predictions from the standard interaction
model.  The continuous monitoring of SN 1993J with VLBI arrays at
centimeter wavelengths have allowed to characterize its expansion and
determine the properties of the shock/CSM interaction
\cite{marcaide97,bartel00}.

No less spectacular than SN 1993J is the case of SN 1987A in the Large
Magellanic Cloud ($D \approx 150$ kpc). SN 1987A faded away very
quickly (a few days) at radio wavelengths, indicating that the shock
front had gone past the presupernova wind, which was a weird behaviour
for a Red Super Giant (RSG) progenitor. When it was clear that SN 1987A went
off while in its BSG phase, it was predicted that once the
shock front would reach the RSG wind of the progenitor
star, we would then witness a rise in flux density. In fact, SN 1987A
reemerged at radio wavalengths in mid 1990, and images taken in 1996
and later on, showed a remarkably circularly symmetric structure, as
displayed by SN 1993J. Recently, SN 1987A was successfully imaged with
e-VLBI (Tingay et al., see these proceedings).

VLBI observations of SN\,1986J (Type IIn) in NGC 891 ($D \approx 9.6$
Mpc) taken approximately 16 yr after its explosion showed that SN
1986J has a distorted, rather than circularly symmetric, shell of
radio emission, indicative of a deformation of the shock front, and
probably due to the collision of the supernova ejecta with an
anisotropic, clumpy medium \cite{pereztorres02}.  Further
multi-wavelength VLBI observations of SN 1986J have shown that this
supernova hosts a compact object in its center \cite{bietenholz04},
tentatively identified with a pulsar, the remnant left after the
explosion of the progenitor star.
SN 1979C (Type II L) in M100 ($D \approx 16.1$ Mpc) is another
remarkable supernova that in the last years has shown an increasing
flux density. This has allowed for VLBI imaging of its fine radio
structure and for a monitoring that has permitted to determine its
deceleration parameter, $m$ ($r \propto t^m$).  SN 2001gd (Type IIb)
and SN 2004et (Type IIP) have also been imaged with VLBI, but the
large distance and relatively quick evolution of SN 2001gd
\cite{pereztorres05} and the very faint event yielded by SN 2004et
\cite{martividal07} resulted in VLBI images which do not resolve
adequately their radio brightness structure.

\section{High-resolution radio imaging of starburst galaxies}

A large fraction of the massive star-formation at both low- and
high-$z$ has taken place in (U)LIRGs. Thus, their implied high
star-formation rates (SFRs) are expected to result in CCSN rates a
couple of orders of magnitude higher than in normal
galaxies. Therefore, a powerful tracer for starburst activity in
(U)LIRGs is the detection of CCSNe, since the SFR is directly related
to the CCSN rate.  However, most SNe occurring in ULIRGs are optically
obscured by large amounts of dust in the nuclear starburst
environment, and have therefore remained undiscovered by (optical) SN
searches.  Fortunately, it is possible to discover these CCSNe through
high-resolution radio observations, as radio emission is free from
extinction effects.  Furthermore, CCSNe are expected, as opposed to
thermonuclear SNe, to become strong radio emitters when the SN ejecta
interact with the circumstellar medium (CSM) that was ejected by the
progenitor star before its explosion as a supernova.  Therefore, if
(U)LIRGs are starburst-dominated, bright radio SNe are expected to
occur and, given its compactness and characteristical radio behaviour,
can be pinpointed with high-resolution, high-sensitivity radio
observations (e.g., SN 2000ft in NGC 7469 \cite{colina01}; SN 2004ip in
IRAS 18293-3413, \cite{pereztorres07}; SN 2008cs in IRAS 17138-1017,
\cite{pereztorres08a}, \cite{kankare08}; supernovae in Arp 299 \cite{neff04},
Arp 220 \cite{smith98,lonsdale06,parra07}, Mrk 273
\cite{bondi05}). However, since (U)LIRGs are likely to have an AGN
contribution, it is mandatory the use of high-sensitivity,
high-resolution radio observations to disentangle the nuclear and
stellar (mainly from young SNe) contributions to the radio emission,
thus probing the mechanisms responsible for the heating of the dust in
their (circum-)nuclear
regions.

One of the most remarkable cases among (U)LIRGs is the prototypical
ULIRG Arp 220. Its central region is composed of two radio nuclei,
each of them hosting a large number of compact components, mostly
identified with young SNe and SNRs
\cite{smith98,lonsdale06,parra07}.  The radio
light curves of many of those SNe appeared to be stable over periods
$\geq$5 yr, indicating their interaction with a dense ISM
\cite{rovilos05}.  The eastern nucleus hosts 20 RSNe, while the
western one hosts 29 RSNe, and its about three times brighter,
suggesting that there might be a differential free-free absorption in
the nuclei \cite{lonsdale06}.  Recently, \cite{parra07} have detected
many of the compact components in Arp 220 at frequencies above 1.6
GHz, and their spectra indicate the presence of both relatively young
SNe along with SNRs. Moreover, the radio supernova rate found by
\cite{parra07} coincides with the expected CCSN rate (4 SN/yr)
as inferred from Arp 220 infrared luminosity. Since all the new RSNe
are very bright, indicating they are Type IIn-like SNe (whose
progenitor stars are very massive), this indicate a top-heavy IMF for
the stars in the nuclei of Arp 220.

\section{e-VLBI imaging of Arp 299-A}

Arp~299 (the merging system formed by IC 694 and NGC 3690) is the
``original'' starburst galaxy (Gehrz et al. 1983) and an obvious
merger system that has been studied extensively at many wavelengths.
An active starburst in Arp~299 is indicated by the high frequency of
recent optically discovered supernovae in the outer regions of the
galaxy.  Since the far infrared luminosity of Arp 299 is $ L_{IR}
\approx 6.5\times 10^{11} L_{\odot} $, the implied CCSN rate is of
$\approx$1.7 SN/yr. Given that 50\% of its total infrared emission
comes from source A (see Fig. \ref{fig,e-EVN}), it is expected that roughly 1 SN/yr will explode
in region A. Therefore, this region is the one that shows most
promises for finding new supernovae. Indeed, Neff et al. (2004) found
a new component in this region, by comparing VLBA observations carried
out in April 2002 and February 2003. 

We proposed e-EVN observations aimed at detecting the radio emission from recently exploded SNe in Arp 299.  Our observations, carried out in April 2008 and December 2008 at 5.0 GHz, resulted in the deepest images ever of Arp 299-A (see Figure \ref{fig,e-EVN}).  We found 26 compact components above 5$\sigma$ rms noise, whose nature can be only explained if they are SNe and/or SNRs, likely embedded in super star clusters, and may challenge the standard scenario
 that directly links far-infrared luminosity to a CCSN and star
 formation rate, since the apparent rate of CCSNe might be much higher than expected.
We leave, however, a detail discussion of this and other issues for future publications.

\begin{figure}[htbp!]
\includegraphics[width=0.8\textwidth]{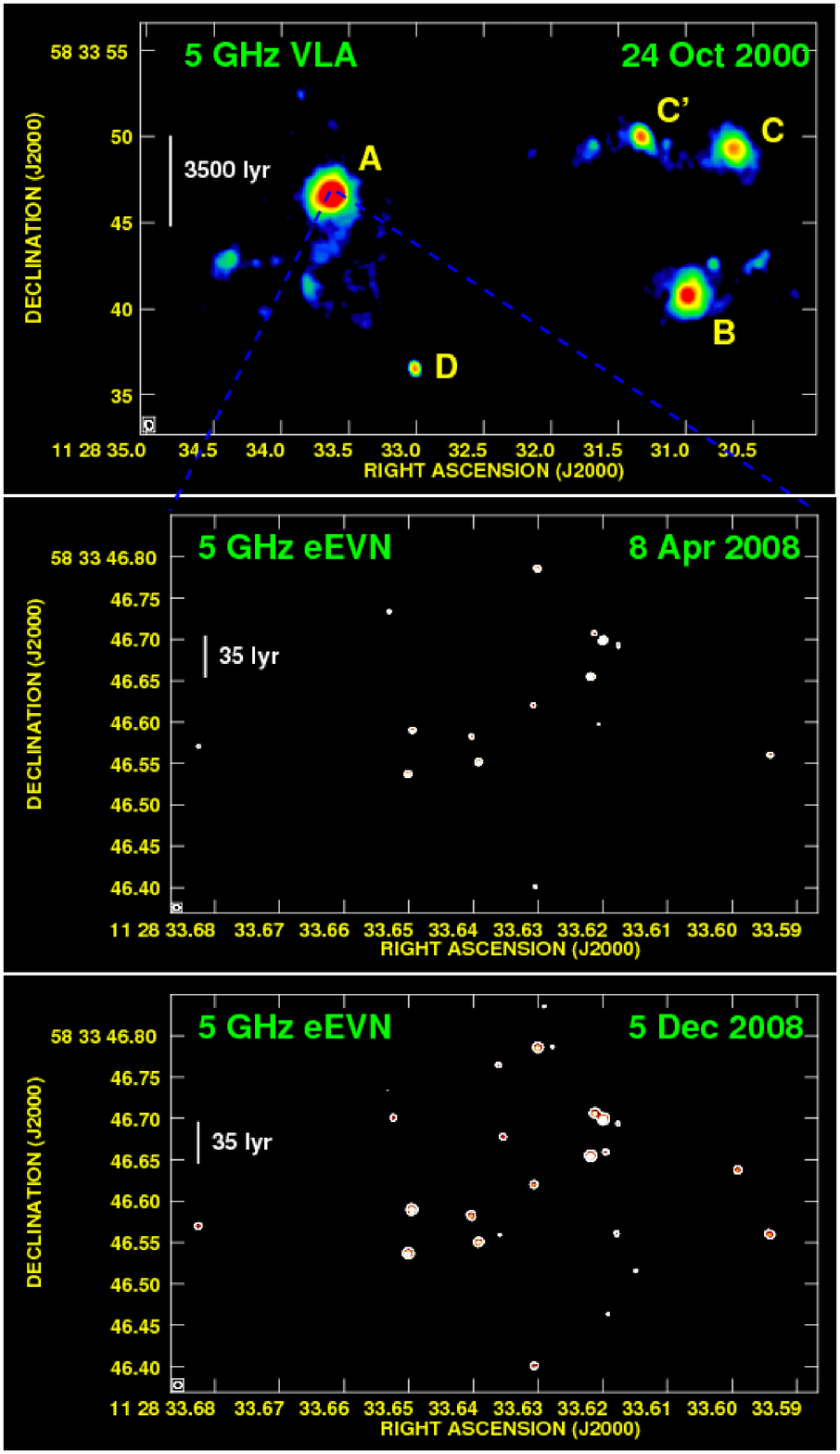}
\caption{ {\it Top panel:} 5 GHz VLA archival observations of Arp 299
  on 24 October 2000, displaying the five brightest knots of radio
  emission in this merging galaxy.  {\it Middle and bottom panels:} 5
  GHz e-EVN observations of the central 500 light years of Arp 299-A on
  8 April 2008 and 5 December 2008.  The off-source root-mean-square
  (r.m.s.) noise level is 39 $\mu$Jy/beam and 25 $\mu$Jy/beam for the
  middle and bottom panels, respectively, and show the existence of 15
  and 26 compact components with a signal-to-noise ratio (s.n.r.)
  equal or larger than five on 8 April 2008 and 5 December 2008,
  respectively.  The size of the FWHM synthesized interferometric beam
  was of (0.6 arcsec $\times$ 0.4 arcsec) for the VLA observations,
  and of (7.3 milliarcsec $\times$ 6.3 milliarcsec) and (8.6
  milliarcsec $\times$ 8.4 milliarcsec) for the e-EVN observations on 8
  April 2008 and 5 December 2008, respectively.}
\label{fig,e-EVN}
\end{figure}

\section{Summary and discussion}

VLBI observations of nearby CCSNe have allowed for a better
understanding of the physics, namely the determination of the
deceleration parameter, interaction between ejecta and presupernova
wind, characterization of the mass loss history and --sometimes-- the
explosion scenario, estimation of magnetic field and energy budget in
fields and particles. However, this wealth of information has been
obtained only for those supernovae that are bright ($L_{\rm peak}
\gsim 1.5 \EE{27}$ \ergshz), long-lasting (radio lifetimes of a few
years at least) and close enough ($D < 20$ Mpc), so that VLBI
observations can adequately resolve and monitor their expansion.  If
we consider that normal galaxies have small CCSN rates ($\lsim 0.01$
SN/yr), and that most CCSNe (Type IIP and Type IIb) have radio peaks
of a few times $10^{26}$\ergshz at most, this explains why so few
radio supernovae have been observed with VLBI in the last 20 yrs,
despite the VLBI arrays increasing their sensitivity.  One importan
way in which e-VLBI may contribute significantly in this field is in
the prompt response that it offers. For example, Type Ib/c SNe,
recently linked to long GRBs, are known to be rapidly evolving
($t_{\rm peak} \approx 10-20$ days) radio supernovae. Currently VLBI
arrays do not offer the needed dynamic scheduling to, e.g., react on a
nearby event, while e-VLBI offers such flexibility and, thanks to its
ability to carry out real-time correlation, allows for a potential
follow-up of the most interesting targets.

The CCSN rate in (U)LIRGs is expected to be at least one or two
orders of magnitude larger than in normal galaxies \cite{condon92},
and hence detections of SNe in (U)LIRGs offer a promising way of
determining the current star formation rate in nearby galaxies.
However, the direct detection of CCSNe in the extreme ambient
densities of the central few hundred pc of (U)LIRGs is extremely
difficult, as the optical and IR emission of supernovae is severely
hampered by the huge amounts of dust present in those regions, and can
at best yield an upper limit to the true CCSN rate.
Fortunately, it is possible to directly probe the star forming
activity in the innermost regions of (U)LIRGs by means of high
angular resolution, high-sensitivity radio searches of CCSNe, as radio does not suffer
from dust obscuration.
Current VLBI (and e-VLBI) arrays are starting to yield astonishing
results, thanks to their few-$\mu$Jy sensitivity and milliarcsec
resolution at cm-wavelengths. In particular, the findings in Arp 220 using global VLBI
 and Arp 299 (using the e-EVN), may challenge the standard scenario
 that directly links far-infrared luminosity to a CCSN and star
 formation rate, which is of much relevance for studies of starburst
 galaxies at high-z.

\begin{small}
\paragraph*{Acknowledgments}
The author acknowledges support by the Spanish
\textsl{Ministerio de Educación y Ciencia (MEC)} through grant AYA
2006-14986-C02-01,  and also by the \textsl{Consejería de Innovación, Ciencia y
Empresa} of  \textsl{Junta de Andaluc\'{\i}a} through grants FQM-1747 and
TIC-126. MAPT is a Ram\'on y Cajal Post Doctoral Research
Fellow funded by the MEC and the \textsl{Consejo Superior de
  Investigaciones Científicas (CSIC)}. 
 The European VLBI Network is a joint facility of European,
Chinese, South African and other radio astronomy institutes funded by
their national research councils.  
e-VLBI developments in Europe are supported by the EC DG-INFSO funded Communication Network Developments project 'EXPReS', Contract No. 02662.
\end{small}

\end{document}